\documentclass[twocolumn,amsmath,aps,fleqn]{revtex4}
\usepackage{graphicx,amssymb}
\begin{document}
\newcommand{\be}{\begin{equation}}
\newcommand{\ee}{\end{equation}}
\newcommand{\bq}{\begin{eqnarray}}
\newcommand{\eq}{\end{eqnarray}}
\newcommand{\bsq}{\begin{subequations}}
\newcommand{\esq}{\end{subequations}}
\newcommand{\bc}{\begin{center}}
\newcommand{\ec}{\end{center}}
\newcommand {\R}{{\mathcal R}}
\newcommand{\al}{\alpha}
\newcommand\lsim{\mathrel{\rlap{\lower4pt\hbox{\hskip1pt$\sim$}}
    \raise1pt\hbox{$<$}}}
\newcommand\gsim{\mathrel{\rlap{\lower4pt\hbox{\hskip1pt$\sim$}}
    \raise1pt\hbox{$>$}}}

\title{Mass inflation in Eddington-inspired Born-Infeld black holes: analytical scaling solutions}

\author{P.P. Avelino}
\email[Electronic address: ]{pedro.avelino@astro.up.pt}
\affiliation{Instituto de Astrof\'{\i}sica e Ci\^encias do Espa{\c c}o, Universidade do Porto, CAUP, Rua das Estrelas, PT4150-762 Porto, Portugal}
\affiliation{Centro de Astrof\'{\i}sica da Universidade do Porto, Rua das Estrelas, PT4150-762 Porto, Portugal}
\affiliation{Departamento de F\'{\i}sica e Astronomia, Faculdade de Ci\^encias, Universidade do Porto, Rua do Campo Alegre 687, PT4169-007 Porto, Portugal}

\date{\today}
\begin{abstract}

We study the inner dynamics of accreting Eddington-inspired Born-Infeld black holes using the homogeneous approximation and taking charge as a surrogate for angular momentum. We show that there is a minimum of the accretion rate below which mass inflation does not occur, and we derive an analytical expression for this threshold as a function of the fundamental scale of the theory, the accretion rate, the mass, and the charge of the black hole. Our result explicitly demonstrates that, no matter how close Eddington-inspired Born-Infeld gravity is to general relativity, there is always a minimum accretion rate below which there is no mass inflation. For larger accretion rates, mass inflation takes place inside the black hole as in general relativity until the extremely rapid density variations bring it to an abrupt end. We derive analytical scaling solutions for the value of the energy density and of the Misner-Sharp mass attained at the end of mass inflation as a function of fundamental scale of the theory, the accretion rate, the mass, and the charge of the black hole, and compare these with the corresponding numerical solutions. We find that, except for unreasonably high accretion rates, our analytical results appear to provide an accurate description of homogeneous mass inflation inside accreting Eddington-inspired Born-Infeld black holes.

\end{abstract}
\maketitle

\section{\label{intr}Introduction}

The formulation of Eddington-inspired Born-Infeld (EiBI) gravity \cite{Banados:2010ix} (see also \cite{Deser:1998rj, Vollick:2003qp,Vollick:2005gc,Vollick:2006qd}) has been inspired by Born-Infeld non-linear electrodynamics \cite{Born:1934gh} and its solution to the problem of the divergent self-energy of point charges. Analogously, the potential avoidance of astrophysical and cosmological singularities has been the main drive behind the development of EiBI gravity. Although, this theory is completely equivalent to Einstein's general relativity in vacuum, significant deviations from general relativity manifest themselves if the energy density or its space-time gradients are sufficiently large \cite{Pani:2011mg,Casanellas:2011kf,Avelino:2012ge,Avelino:2012qe,Sham:2012qi,Harko:2013wka,Sotani:2014goa,Sotani:2014xoa}. This is particularly true in the extreme environments attained in the early universe \cite{Avelino:2012ue,Scargill:2012kg,Cho:2013pea,Bouhmadi-Lopez:2014jfa,Jimenez:2014fla,Harko:2014nya,Cho:2014jta,Cho:2015yua,Jimenez:2015jqa} and inside black holes \cite{Olmo:2013gqa,Harko:2013aya,Fernandes:2014bka,Wei:2014dka,Sotani:2014lua,Shaikh:2015oha,Jana:2015cha,Olmo:2015dba,Bazeia:2015uia,Olmo:2015bya}, which may, if certain conditions are verified, be singularity free (see, however, \cite{EscamillaRivera:2012vz,Pani:2012qd,Sham:2013sya,Yang:2013hsa,Kim:2013nna}).

An exponential growth of the Misner-Sharp mass, known as mass inflation, has been shown arise as a consequence of the relativistic counterstreaming between ingoing and outgoing streams inside charged Reissner-Nordstr$\ddot{\rm o}$m charged black holes and rotating Kerr black holes in the context of General Relativity \cite{Poisson:1989zz,Hod:1998gy,Ori:1991zz,Hansen:2005am,Hamilton:2008zz} (see also  \cite{Avelino:2009vv,Avelino:2011ee,Hansen:2014rua,Avelino:2014aea,Hansen:2015dxa} for a number of other studies in the context of modified gravity). The role of mass inflation on the inner dynamics of EiBI black holes has been investigated for the first time in \cite{Avelino:2015fve}. There it has been found, using numerical simulations of accreting spherically symmetric charged black holes in the homogeneous approximation, that there is a minimum accretion rate for mass inflation to occur in the context of EiBI gravity. In this paper we extend these results by performing a detailed analytical study of mass inflation inside EiBI black holes.

Throughout this paper we shall use fundamental units with $c=G=1$ and a metric signature $(-,+,+,+)$. The Einstein summation convention will be used when a greek index, taking the values $0,...,3$, appears twice in a single term (the exception will be the greek indices $\theta$ and $\phi$ which will denote the polar and azimuthal angles, respectively). 

\section{EiBI gravity \label{sec2}}

EiBI gravity is described by the action
\be
S=\frac{2}{\kappa}\int d^{4}x\left[\sqrt{\left|g_{\mu\nu}+\kappa R_{\mu\nu}\right|}-\lambda\sqrt{|g|}\right]+S_M\,,\label{eq:EddingtonBornInfeld Action}
\ee
and it is based on the Palatini formulation which treats the metric and the connection as independent fields.
Here, $g_{\mu\nu}$ are the components of the metric, $g$ is the determinant of $g_{\mu\nu}$, $R_{\mu\nu}$ is the symmetric Ricci tensor build from the connection  $\Gamma$, $S_M$ is the standard action associated with the matter fields, and $\kappa$ is the only additional parameter of the theory with respect to general relativity (see \cite{Pani:2011mg,Avelino:2012ge,Avelino:2012qe} for tight constraints on the value of $\kappa$). 

Therefore, the equations of motion may be derived by varying the action with respect to the connection and the metric. These are given respectively by
\bq
q_{\mu\nu}&=&g_{\mu\nu}+\kappa R_{\mu\nu}\,,\label{eq:ConnectionEquationOfMotion}\\
\sqrt{|q|}q^{\mu\nu}&=&\lambda\sqrt{|g|}g^{\mu\nu}-{\bar \kappa}\sqrt{|g|}T^{\mu\nu}\,,\label{eq:MetricEquationOfMotion}
\eq
where $T^{\mu \nu}$ are the components of the energy-momentum tensor, $q_{\mu\nu}$ is an auxiliary metric related to the original connection by 
\be
\Gamma^{\gamma}_{\mu\nu} = {1 \over 2} q^{\gamma\zeta}(q_{\zeta\mu,\nu} + q_{\zeta\nu,\mu}- q_{\mu\nu,\zeta})\,, \label{connection}
\ee
$q^{\mu\nu}$ is the inverse of $q_{\mu\nu}$, ${\bar \kappa}=8\pi \kappa$ and a comma represents a partial derivative. Without loss of generality we set $\lambda=1$. Although the changes associated with a different value of $\lambda$ can  be incorporated into the energy-momentum tensor, in this paper we shall not consider them since we will only be dealing with asymptotically flat solutions.

Combining Eqs.~(\ref{eq:ConnectionEquationOfMotion}) and (\ref{eq:MetricEquationOfMotion}) one obtains the second-order field equations
\be
{{\mathcal G}^\mu}_\nu \equiv {{{\mathcal R}}^\mu}_\nu -\frac12 {\mathcal R} {\delta^\mu}_\nu  =8\pi {{\mathcal T}^{\mu}}_{\nu}\,,\label{eq:EquationOfMotionComb1}
\ee
with 
\bq
{{{\mathcal R}}^\mu}_\nu &\equiv& q^{\mu \zeta} R_{\zeta \nu} =8\pi{\Theta^{\mu}}_{\nu}\,,\label{eq:EquationOfMotionComb}\\
{{\mathcal T}^{\mu}}_{\nu} &\equiv& {\Theta^{\mu}}_{\nu}-\frac12\Theta {\delta^\mu}_\nu\,,\\
{\Theta^{\mu}}_{\nu}&\equiv&\frac{1}{{\bar \kappa}}\left(1-\sqrt{{\left|\frac{g}{q}\right|}}\right){\delta^\mu}_\nu+\sqrt{{\left|\frac{g}{q}\right|}}{T^{\mu}}_{\nu} \,,\label{eq:Thetamunu}\\
\Theta &\equiv&  {\Theta^{\mu}}_{\mu}\,.\label{eq:Theta}
\eq

For $|{\bar \kappa}| \rho \ll 1$, and if the density field is smooth enough, the components of the physical ($g$) and auxiliary ($q$) metrics are approximately equal and ${{\mathcal T}^{\mu}}_{\nu} \sim {T^{\mu}}_{\nu}$. Consequently, in vacuum the EiBI theory of gravity is indistinguishable from general relativity.

\section{\label{sec3} Spherically symmetric homogeneous approximation}

Rotating and charged black holes have similar geometries. Hence, in this paper we will use charge as a surrogate for angular momentum, considering accreting spherically symmetric charged black holes. We shall also consider the homogeneous approximation in the computation of the black hole's interior structure, thus assuming that all relevant quantities can be written as a function of a radial (timelike) coordinate alone. This approximation has been shown to provide an accurate description of some of the most important aspects of mass inflation (see, e.g., \cite{Hansen:2005am,Avelino:2009vv,Avelino:2011ee}). 

In the homogeneous approximation the spherically symmetric physical ($g$) and auxiliary ($q$) line elements may be written, respectively,  as
\bq
ds_g^2&=&g_{tt}(r) dt^2+g_{rr}(r) dr^2+r^2(d\theta^2+\sin^2\theta d\phi^2)\,, \label{le1}\\
ds_q^2&=&A(r) dt^2+B(r) dr^2+H^2(r)(d\theta^2+\sin^2\theta d\phi^2)\,, \label{le2}
\eq
where $g_{tt}$, $g_{rr}$, $A\equiv q_{tt}$, $B \equiv q_{rr}$, and $H^2 \equiv q_{\theta \theta}$ are all functions of $r$ alone. 

The non-zero components of the energy-momentum tensor of the electric field corresponding to a constant charge $Q$ are given by
\bq
{^eT^r}_r&=&-\rho_e\,,  {^eT^t}_t=w_{e\parallel} \rho_e \,, \\
{^eT^\theta}_\theta&=&{^eT^\phi}_\phi=w_{e\perp} \rho_e\,,
\eq
with
\be
w_{e\parallel}=-1\,, \qquad w_{e\perp}=1\,,
\ee
and
\be
\rho_e=\frac{Q^2}{8\pi r^4}\,. 
\ee

The non-zero components of the most general fluid energy-momentum tensor consistent with spherical symmetry and  the homogeneous approximation are given by
\bq
{^fT^r}_r&=&-\rho_f\,,  {^fT^t}_t=p_{f\parallel}=w_{f\parallel} \rho_f \,, \\
{^fT^\theta}_\theta&=&{^fT^\phi}_\phi=p_{f\perp}=w_{f\perp} \rho_f\,,
\eq
where $\rho_f$, $p_{f\parallel}$, and $p_{f\perp}$ are the fluid's density, radial pressure and transverse pressure, respectively.

Energy-momentum conservation of the accreting fluid implies that
\be
\frac{\rho_f'}{\rho_f}=- \frac{1+w_{f\parallel}}{2}\frac{g_{tt}'}{g_{tt}}-\frac{2(1+w_{f\perp})}{r}\label{rhoeq}\,,
\ee
where a prime represents a derivative with respect to the timelike coordinate $r$. Integrating Eq.~(\ref{rhoeq}) with respect to $r$ one obtains
\be
\rho_f =  \rho_{fi} \left(\frac{g_{tti}}{g_{tt}}\right)^{(1+w_{f\parallel})/2} \left(\frac{r_i}{r}\right)^{2(1+w_{f\perp})}\,, \label{denf}
\ee
with the subscript $i$ meaning that the physical quantities are evaluated at some initial radius $r_i$.

The total energy-momentum tensor
\be
{T^\mu}_\nu={^fT^\mu}_\nu+{^eT^\mu}_\nu\,,
\ee
is just the sum of the fluid and electromagnetic parts, which are assumed to be separately conserved. Consistently with the above notation, we shall also write
\bq
{T^r}_r&=&-\rho\,,  {T^t}_t=p_{\parallel}=w_\parallel \rho \,, \\
{T^\theta}_\theta&=&{T^\phi}_\phi=p_{\perp}=w_\perp \rho\,,
\eq
with $\rho$, $p_{\parallel}$, and $p_{\perp}$ being the total density, radial pressure and transverse pressure, respectively.

The following relations between the components of the physical and auxiliary metrics can be computed using Eqs.~(\ref{eq:ConnectionEquationOfMotion}) and (\ref{eq:MetricEquationOfMotion})
\bq
A&=&g_{tt} \frac{(1+{\bar \kappa} \rho)^{1/2} (1-{\bar \kappa} w_\perp \rho)}{(1-{\bar \kappa} w_\parallel \rho)^{1/2}}\,, \\ 
B&=&g_{rr} \frac{(1-{\bar \kappa} w_\parallel \rho)^{1/2} (1-{\bar \kappa} w_\perp \rho)}{(1+{\bar \kappa} \rho)^{1/2}}\,,\\
H&=&r(1+{\bar \kappa}\rho)^{1/4}(1+{\bar \kappa} w_\parallel\rho)^{1/4},
\eq
and they imply that
\be
\sqrt{{\left|\frac{g}{q}\right|}}=\left(1+{\bar \kappa} \rho\right)^{-1/2} \left(1-{\bar \kappa} w_\parallel \rho\right)^{-1/2} \left(1-{\bar \kappa} w_\perp \rho\right)^{-1}\,.
\ee

If $\rho_f=0$ then the standard Reissner-Nordstr$\ddot{\rm o}$m solution with
\bq
A&=&-\left(1-\frac{2M}{r}+\frac{Q^2}{r^2}\right) \\ 
B&=&-\frac{1}{A}\, \\ 
H&=&r\,,
\eq
is an excellent approximation both in the $\kappa \to 0$ limit or for sufficiently large values of $r$. For $\kappa=0$ the outer ($r_+$) and inner ($r_-$) horizons of the black hole are located at
\be
r_{\pm}=\left(M \pm {\sqrt{M^2-Q^2}}\right)\,.
\ee

\section{\label{sec4} Mass inflation: analytical solutions}

It has been shown that, if mass inflation occurs, the relativistic counterstreaming between ingoing and outgoing streams drives $w_{f\parallel}$ towards unity in the mass inflation region \cite{Hamilton:2008zz}. Hence, here we shall now investigate the mass inflation regime with $w_\parallel \sim w_{f\parallel}\sim1$ and $|{\bar \kappa}| \rho \ll 1$.
Under these conditions the relations 
\be
A=g_{tt}\,, \quad B=g_{rr}\,, \quad H=r \,,
\ee 
approximately hold. We shall also assume that mass inflation takes place for 
\be
r \sim r_-\,, \label{r}
\ee
which has been shown to be a good approximation for reasonable (not too large) values of the accretion rate (see, e.g., \cite{Hansen:2005am,Hamilton:2008zz,Avelino:2009vv,Avelino:2011ee}). 

Given that during mass inflation regime $\rho_f$ becomes much larger than $\rho_e$ (so that $\rho \sim \rho_f$), in this period $H$ is given approximately by
\be
H \sim r (1+{\bar \kappa} \rho)^{1/4} (1- {\bar \kappa} \rho)^{1/4} \sim r_- \left(1- \left(\frac{{\bar \kappa} \rho}{2}\right)^2 \right)\,,\label{Happ}
\ee
where the last approximation in Eq.~(\ref{Happ}) is valid up to first order in $|{\bar \kappa}| \rho$. The first and second derivatives of H satisfy
\bq
H'&\sim&1-r \frac{{\bar \kappa}^2}{2} \rho \rho' \sim 1-r _- \frac{{\bar \kappa}^2}{2} \rho \rho' \,, \label{hprime}\\
H''&\sim&-r\frac{{\bar \kappa}^2}{2} \left(\rho'^2+\rho \rho'' \right) \sim -r_- ({\bar \kappa} \rho')^2   \label{hdprime}\,.
\eq
Note that the last approximation in Eq.~(\ref{hdprime}) takes into account that $\rho'/\rho \sim {\rm const}$, during mass inflation \cite{Avelino:2012ue}.

The $tt$ and $rr$ components of Eq.~(\ref{eq:EquationOfMotionComb1}) are given by
\bq
-\frac{H'}{H}\frac{B'}{B}-\frac{B}{H^2}-\left(\frac{H'}{H}\right)^2+2\frac{H''}{H}&=&8\pi B {{\mathcal T}^t}_t\,, \label{Beq}\\
-\frac{H'}{H}\frac{A'}{A}+\frac{B}{H^2}-\left(\frac{H'}{H}\right)^2&=&8\pi B{{\mathcal T}^r}_r \label{Aeq}\,.
\eq
Mass inflation takes place for $r_- {\bar \kappa}^2 \rho |\rho'| \ll 1$, so that $H' \sim 1$ (in this period the conditions $H \sim r \sim r_-$, $|{\bar \kappa}| \rho \ll 1$, $A \sim  g_{tt}$, and $B \sim  g_{rr}$ are also satisfied). In the mass inflation regime the values of $|g_{rr}|$ and $|g_{tt}|$ become tiny and Eqs.~(\ref{Beq}) and (\ref{Aeq}) are given approximately by
\bq
\frac{g_{rr}'}{g_{rr}}&\sim&-8\pi r_- \rho g_{rr} \label{grreq}\,,\\
\frac{g_{tt}'}{g_{tt}}&\sim&-8\pi r_- \rho g_{rr}  \label{gtteq}\,,
\eq
respectively. The last term on the left-hand side of Eq.~(\ref{Beq}) may be neglected, compared to the source term on the right-hand side of the same equation, for $r_- {\bar \kappa}^2 \rho |\rho'| \ll 1$ (note that $\rho'/\rho \sim -g_{rr}'/g_{rr} \sim -g_{rr}'/g_{rr} \sim {\rm const}$ during mass inflation).

Subtracting Eqs.~(\ref{grreq}) and (\ref{gtteq}), and integrating the result with respect to $r$, one obtains
\be
\left.\frac{g_{rr}}{g_{tt}}\right|_{\rm [MI]} \sim {\rm const} \,, \label{grrogttMI}
\ee
where the label $[\rm MI]$ indicates that the corresponding quantity is to be evaluated during mass inflation.
Eq.~(\ref{grrogttMI}) implies that the value of $g_{rr}/g_{tt}$ is approximately the same at the start and at the end of mass inflation, or, equivalently, that
\be
\left.\frac{g_{rr}}{g_{tt}}\right|_{\rm [start]} \sim \left.\frac{g_{rr}}{g_{tt}}\right|_{\rm [end]}\,. \label{grrogttstart}
\ee

On the other hand, 
\be
g_{tt[\rm start]} \sim - g_{rr[\rm start]}^{-1}\,,  \label{gttstart}
\ee
since  $g_{rr} g_{tt}  \sim -1$ while $\rho_f$ remains smaller than $\rho_e$ before the start of mass inflation. Eqs. (\ref{grrogttstart}) and (\ref{gttstart}) imply that
\be
\left.\frac{g_{rr}}{g_{tt}}\right|_{\rm [MI]}= \left. \frac{g_{rr}}{g_{tt}} \right|_{\rm [start]} \sim -g_{rr[\rm start]}^2 \sim -g_{tt[\rm start]}^{-2}\,. \label{grrogttMI1}
\ee

Mass inflation starts when the energy density of the accreting fluid begins to dominate over the energy density associated to the electric field. For the purpose of finding analytical solutions, we shall assume that mass inflation starts when $\rho_f=\alpha \rho_e$, or equivalently, when
\be
\rho_{f[\rm start]}=\alpha\frac{Q^2}{8\pi r_-^4} \label{massinfstart}\,,
\ee
where $\alpha$ is of order unity. Taking into account that (see Eq.~(\ref{denf}) for $w_{f\parallel} \sim 1$ and $r \sim r_-$)
\be
\rho_{f[\rm start]} \sim  \rho_{fi} \frac{g_{tti}}{g_{tt[\rm start]}} \left(\frac{r_i}{r_-}\right)^{2(1+w_{f\perp})}\,, \label{den1}
\ee
and using Eqs.~(\ref{grrogttMI1}) and (\ref{massinfstart}), one obtains
\be
\left.\frac{g_{rr}}{g_{tt}}\right|_{\rm [MI]}\sim-\frac{\alpha^2 Q^4}{64 \pi^2 \rho_{fi}^2 \, g_{tti}^2 \, r_-^{4(1-w_{f\perp})} r_i^{4(1+w_{f\perp})}} \,. \label{grrogtt}
\ee

Mass inflation ends when $H'$ starts deviating significantly from unity. In order to find analytical solutions we shall assume that mass inflation ends when 
\be
r_-\frac{{\bar \kappa}^2}{2} \rho |\rho'| = \beta\,, 
\ee
where
\bq
\frac{\rho'}{\rho} &\sim& -\frac{g_{tt}'}{g_{tt}} \sim  8\pi r_-\rho g_{rr}  \nonumber \\ 
&\sim& 8 \pi r_-\rho_{fi} g_{tti} \left.\frac{g_{rr}}{g_{tt}}\right|_{\rm [MI]}  \left(\frac{r_i}{r_-}\right)^{2(1+w_{f\perp})}\,, \label{rhoprime1}
\eq
is roughly constant during mass inflation and $\beta$ is of order unity. Using Eqs.~(\ref{den1}), (\ref{grrogtt}) and (\ref{rhoprime1}) one finds that the maximum density attained at the end of mass inflation is given by
\be
\rho_{\rm [end]}\sim\frac{\beta^{1/2}}{2 \pi^{1/2}\alpha} \frac{g_{tti}^{1/2} r_-^{2-w_{f\perp}}r_i^{1+w_{f\perp}}  }{ Q^2 } \frac{\rho_{fi}^{1/2}}{|\kappa|}\,. \label{rhoend}
\ee
Hence, there is a minimum accretion threshold, parameterized by $\rho_{fi}$, for mass inflation to occur. If
\be
\rho_{\rm [end]} <  \rho_{f{\rm[start]}} = \alpha \frac{Q^2}{8\pi r_-^4}\,,
\ee
or, equivalently,
\be
\frac{\rho_{fi}^{1/2}}{|\kappa|} < \frac{\alpha^2}{4 \pi^{1/2}\beta^{1/2}}   \frac{Q^4}{g_{tti}^{1/2}r_-^{6-w_{f\perp}} r_i^{1+w_{f\perp}} }\,, \label{treshold}
\ee
mass inflation does not happen at all. Therefore, our characterization of the mass inflation regime inside EiBI black holes is only valid above this threshold.

\begin{figure}
\includegraphics[width=3.4in]{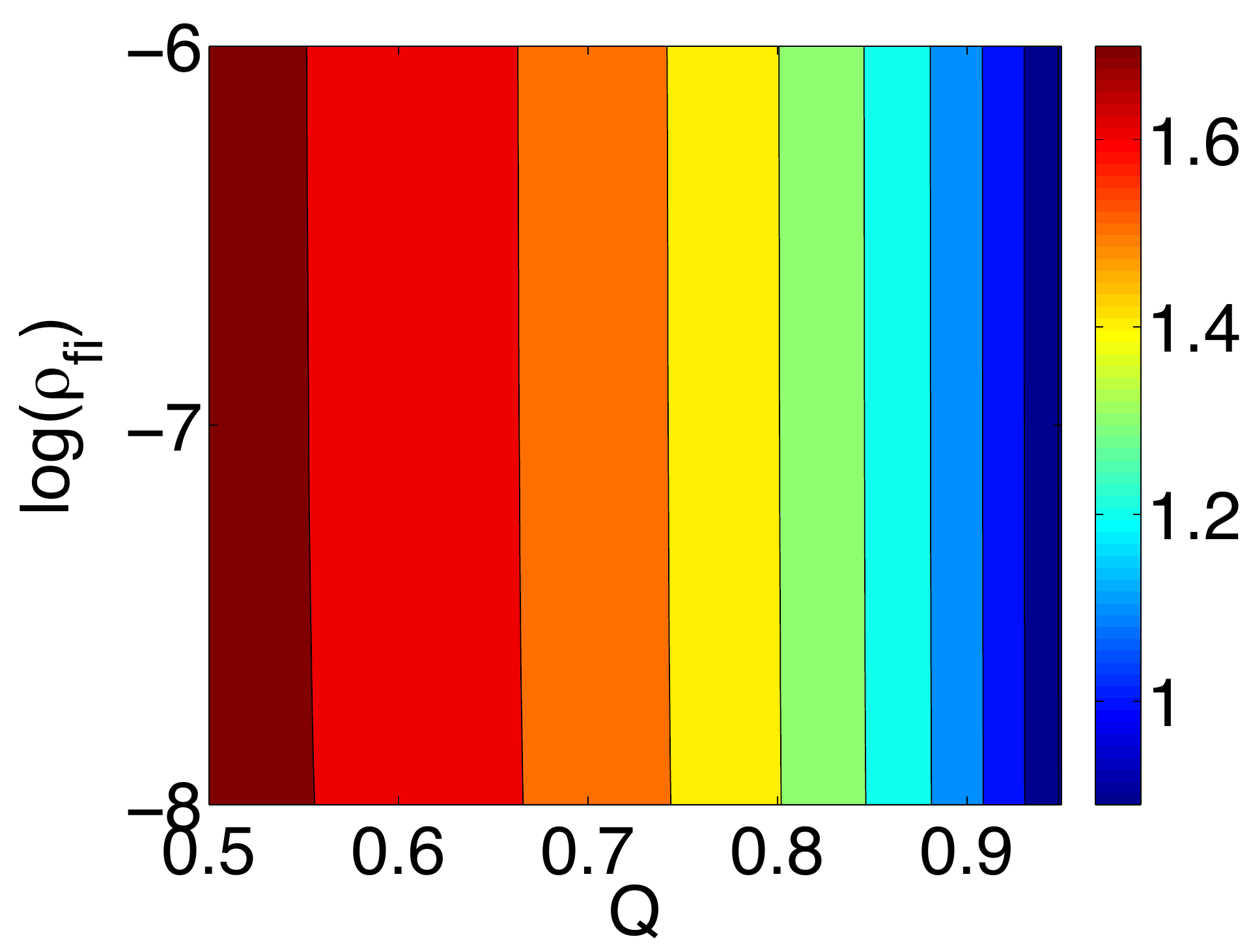}
\includegraphics[width=3.4in]{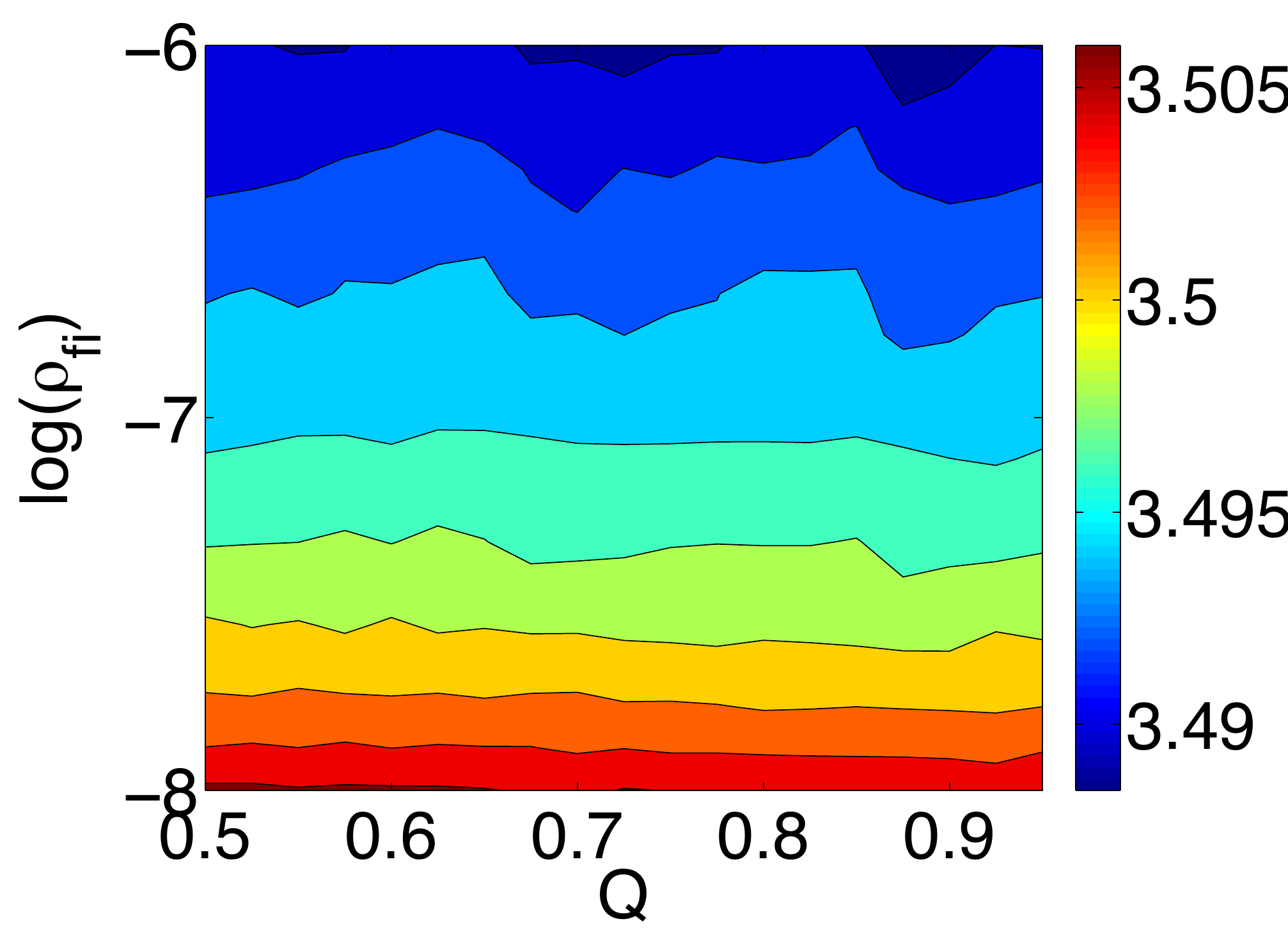}
\caption{The values of $\alpha$ (top panel) and $\beta$ (bottom panel) for $\kappa = {\bar k}/ (8 \pi) = 10^{-40}$, considering $Q$ and $\rho_{fi}$ in the intervals $[0.5,0.95]$ and $[10^{-8},10^{-6}]$, respectively.}
\label{Fig1}
\end{figure}

The total effective mass inside a sphere of radius $r$ is given by the Misner-Sharp mass
\be
M_{\rm M-S}=\frac{r}{2}\left(1+\frac{q^2}{r^2}-\frac{1}{g_{rr}}\right)\,.
\ee
The maximum of the Misner-Sharp mass, attained at the end of mass inflation, is
\be
M_{\rm M-S[end]} \sim \frac{r_-}{2}\left(1+\frac{q^2}{r_-^2}-\frac{1}{g_{rr[{\rm end}]}}\right) \label{MMSend}\,.
\ee
Here,
\be
\frac{1}{g_{rr[\rm end]}}\sim \frac{1}{\left.\frac{g_{rr}}{g_{tt}}\right|_{\rm [end]}g_{tt[\rm end]}}  \sim 
\frac{1}{\left.\frac{g_{rr}}{g_{tt}}\right|_{\rm [MI]}g_{tt[\rm end]}}\,,
\ee
where
\be
\frac{1}{g_{tt[\rm end]}}\sim \frac{1}{g_{tti}}\frac{\rho_{f[\rm end]}}{\rho_{fi}} \left(\frac{r_i}{r_-}\right)^{-2(1+w_{f\perp})}\,,
\ee
with $(g_{rr}/g_{tt})_{\rm [MI]}$ and $\rho_{[\rm end]}$ given by Eqs.~(\ref{grrogtt}) and Eq.~(\ref{rhoend}), respectively. Hence, one finally finds that
\bq
M_{\rm M-S[end]} &\sim&  -\frac{r_-}{2 g_{rr[{\rm end}]}} \sim \frac{16 \pi^{3/2}\beta^{1/2}}{\alpha^3}  \times \nonumber\\
&\times& \frac{g_{tti}^{3/2} r_-^{9-3w_{f\perp}} r_i^{3+3w_{f\perp}}}{Q^6} \frac{\rho_{fi}^{3/2}}{|\kappa|} \label{MMSend}\,.
\eq

\section{Mass inflation: numerical verification}

We have computed numerically the value of $\rho_{\rm [end]}$ and $M_{\rm M-S[end]}$, as a function of $\rho_{fi}$ and $Q$, considering initial conditions with $r_i=0.95 r_-$. For $|{\bar \kappa}| \rho_{fi} \ll 1$ the inner structure of the black hole around $r=r_i$ is close to that of an ordinary charged Reissner-Nordstr$\ddot{\rm o}$m black hole in general relativity and, consequently, we also assume that
\bq
F(r_i)&=&\left(1-\frac{2M}{r_i}+\frac{Q^2}{r_i^2}\right)\,,\\
G(r_i)&=&1\,,\\
H(r_i)&=&r_i\,.
\eq
We combined our numerical results with the analytical scaling solutions, given by Eqs.~(\ref{rhoend}) and (\ref{MMSend}), to estimate the values of $\alpha$ (Fig.~1, top panel) and $\beta$ (Fig.~1, bottom panel) for $\kappa = {\bar k}/ (8 \pi) = 10^{-40}$, considering $Q$ and $\rho_{fi}$ in the intervals $[0.5,0.95]$ and $[10^{-8},10^{-6}]$, respectively. The results shown in Fig.~1 confirm that Eqs.~(\ref{rhoend}) and (\ref{MMSend}) provide an accurate estimate of $\rho_{\rm [end]}$ and $M_{\rm M-S[end]}$ with $\alpha \sim 1.5$ and $\beta \sim 3.5$ in a regime where the accretion rate, parameterized by $\rho_{fi}$, is small, but the ratio $\rho_{fi}^{1/2}/|\kappa|$ is large enough for mass inflation to take place. The numerical results also show that the value of $\alpha$ appears to be a (slowly) growing function of $Q$.

\section{\label{conc} Conclusions}

In this paper we investigated the dynamics of mass inflation inside accreting EiBI black holes using the homogeneous approximation and taking charge as a surrogate for angular momentum. We have shown that there is a minimum accretion rate below which mass inflation does not occur, and we computed analytically this threshold as a function of the fundamental scale of the theory, the accretion rate, the mass, and the charge of the black hole. Our results imply that mass inflation does not happen for sufficiently low accretion rates, independently of how close EiBI gravity is to general relativity. We have also shown that mass inflation inside EiBI black holes, if it occurs, is brought to an end at an energy density much smaller than the fundamental energy density of the theory. We computed the analytical scaling solutions for the energy density and the Misner-Sharp mass at the end of mass inflation, showing that they  provide a good approximation to the corresponding numerical results.

\begin{acknowledgments}

The author thanks Andrew Hamilton, Carlos Herdeiro, and Diego Rubiera-Garcia for enlightening discussions on the subject of mass inflation and for various useful comments and suggestions on this manuscript. This work was supported by Funda{\c c}\~ao para a Ci\^encia e a Tecnologia (FCT) through the Investigador FCT contract of reference IF/00863/2012 and POPH/FSE (EC) by FEDER funding through the program "Programa Operacional de Factores de Competitividade - COMPETE. Funding of this work was also provided by the FCT grant UID/FIS/04434/2013
\end{acknowledgments}


\bibliography{EiBI-MI}

\end{document}